\documentclass{aastex}         
\usepackage{spr-astr-addons}

\begin{document}

\title{Tidal Love numbers of neutron stars in Rastall gravity}


\author{Liang Meng\altaffilmark{}} 
\and 
\author{Dao-Jun Liu\altaffilmark{*}}
\affil{Center for astrophysics and department of Physics, Shanghai Normal University, Shanghai 200234, China}
\email{djliu@shnu.edu.cn} 

\altaffiltext{*}{djliu@shnu.edu.cn}

\begin{abstract}
    Gravitational-wave measurements of the tidal deformability of neutron stars could reveal important information regarding their internal structure, the equation of state of high-dense nuclear matter and gravity in strong field regime. In this work, we extend the relativistic theory of the tidal deformability of neutron stars to Rastall gravity. Both the electric-type and magnetic-type quadrupole tidal Love numbers are calculated for neutron stars in the polytrope model.  It is found that neutron star's tidal Love numbers in Rastall gravity is significantly smaller than those in general relativity. Our results provide new evidence of the degeneracy between the modification of gravity and the equation of state of nuclear matter in neutron stars. 
\end{abstract}


\section{Introduction}\label{s:introduction}

The first observed gravitational waves from a binary neutron star coalescence, GW170817 \citep{LIGOScientific:2017vwq}, have been used to constrain various properties of neutron stars \citep{Raithel:2018ncd,De:2018uhw}. One of the most elegant and straightforward constraints comes from the measurement of the tidal Love numbers (TLNs) of the merging neutron stars. The TLNs characterize the response (deformability) of a body to an external tidal force \citep{poisson-2014}. 
It is pointed out \citep{Flanagan:2007ix} that the early phase of the inspiral depends cleanly on the TLNs, which are dependent on the equation of state (EOS) of the dense matter in the neutron stars. 

However, the TLNs encode information  about  not only the internal structure of the star but also the property of gravity in the strong field regime. Although there exist universal relations, such as so-called I–Love–Q relations, among certain stellar observables (see \citep{YAGI20171} for a review), it is still difficult to determine the EOS from gravitational wave observations in a theory of gravity independent way. Therefore, it will be interesting to quantify the effects from gravity on the TLNs. The theory of TLNs in general relativity (GR) has been developed elegantly \citep{Hinderer:2007mb,Binnington:2009bb,Damour:2009vw}.  GR passed almost all the precision tests so far with flying colors, but  plenty of alternative theories of gravity have been proposed in the literature (see, e.g. \citep{Clifton:2011jh} for a comprehensive review). 
Stellar structure models in modified theories of gravity have also been vastly investigated, for a review, see \citep{Olmo2020}. 
The TLNs of neutron stars, black holes and other compact objects in alternative theories of gravity have been studied  in a few cases in the past years, see, for example, Refs.\citep{Cardoso2017,Yagi:2013awa,Yazadjiev:2018xxk,Silva:2020acr}.

Recently,  Rastall gravity, which was proposed nearly fifty years ago \citep{Rastall:1972swe}, has regained lots of interest and even debates in literature \citep{Visser:2017gpz,Darabi:2017coc}. Static and spherically symmetric neutron star solutions in Rastall gravity has been obtained for dense matter with different equations of state \citep{Oliveira2015,Xi:2020lbt}.  Rastall gravity gives up the usual energy-momentum conservation laws  for the matter sources of gravity, which have only checked in the flat spacetime,  but suggests that the energy-momentum tensor of matter has a nonzero divergence in the curved spacetime. Note that, although Rastall gravity was originally proposed as a non-Lagrangian theory of gravity, there are some works in the recent literature  where different possibilities have been investigated that lead to a Lagrangian that describes the Rastall gravity ( e.g., \citep{PhysRevD.95.101501,DeMoraes:2019mef,Shabani:2020wja}).

It is of interest to investigate the effect of the violation of the classical energy-momentum conservation on the tidal deformability of compact stars. Therefore, in the present paper, we want to calculate a neutron star's tidal Love number in the fully relativistic polytropic models under the framework of Rastall gravity. Throughout the paper, the geometric units $G=c=1$ are used.

\section{Neutron stars in Rastall gravity}\label{sec:NS}

To study the tidal deformability of neutron stars, we need first of all to obtain a neutron star solution.   According to the Rastall's proposal,  the Einstein equations are directly modified to be 
\begin{equation}
    R_{\mu\nu} = 8 \pi  \left(T_{\mu\nu} - \frac{1+\eta}{1+2\eta} g_{\mu\nu}\frac{T}{2}  \right),
\end{equation}
where $R_{\mu\nu}$ and  $g_{\mu\nu}$  are respectively the  Ricci tensor and metric tensor of spacetime.  $T_{\mu\nu}$ and $T \equiv T_{\lambda}^{\lambda}$ denote the energy-momentum tensor and its trace of the matter. Clearly, the parameter $\eta$ describes  the deviation of the theory from GR and when it is set to be zero, GR is recovered. 

Let us consider a static and spherical star solution and the metric of spacetime is described by
\begin{equation}\label{eq:metric}
    ds^2= g_{\mu\nu}^{(0)}dx^{\mu}dx^{\nu}= -e^{2\alpha}dt^2 + e^{2\beta} dr^2+r^2d\Omega^2,
\end{equation}

where the metric functions $\alpha$ and $\beta$ are both functions of radial coordinate $r$ and $d\Omega^2 = d\theta^2+\sin^2\theta d\phi^2$ is the metric on a unit two-sphere.
The matter content of the star is modeled by a perfect fluid, the energy-momentum tensor of which can be written as 
\begin{equation}\label{eq:EMT}
    T_{\mu\nu} = (\rho + p )U_{\mu}U_{\nu} + p g_{\mu\nu}^{(0)},
\end{equation}
where the energy desity $\rho$ and pressure $p$ will be functions of $r$ alone, and $U_{\mu}$ is the 4-velocity with normalization $U^{\mu}U_{\mu} = -1 $. Since we seek static solutions, we can take $U_{\mu} = (e^{\alpha}, 0,0,0)$. Therefore, 
\begin{equation}
    T^{\nu}_{\mu}=\mathrm{diag}(-\rho,p,p,p), \quad T= 3p-\rho.
\end{equation}
It is convenient to introduce mass function $m=m(r)$ that represents the mass within the radius $r$
\begin{equation}
    e^{2\beta} = \left(1-\frac{2 m}{r}\right)^{-1},
\end{equation}
then, the structure of the star is governed by the following differential equations: 
\begin{equation}
    \frac{d m}{d r} = \frac{2 \pi  r^2 [2\rho +3 \eta(\rho + p) ]}{1+2 \eta},\label{eq:TOVm}
\end{equation}
\begin{equation}\label{eq:TOVp}
    \frac{d p}{d r} = -\frac{c_s^2 (p+\rho) \left[(4 \eta +2) m+4 \pi  r^3 ((\eta +2) p+\eta  \rho)\right]}{r (r-2 m) [(\eta +2) c_s^2+\eta ]},
\end{equation}
\begin{equation}\label{eq:TOValpha}
    \frac{d\alpha}{d r } = e^{2\beta}\left[\frac{m}{r^2}+2\pi r \frac{2p+\eta(\rho+p)}{1+2\eta}\right],
\end{equation}
where $ c_s^2=\frac{\partial p}{\partial \rho}$ is the speed of sound in the fluid. Obviously, when $\eta =0$, the above equations reduce to the well-known Tolman-Oppenheimer-Volkoff (TOV) equations. The above  equations are not closed. To determine the internal structure of the star, we need take a form of the equation of state for the matter in the star $p=p(\rho)$.
Let us assume a simple pure neutron star with nucleon-nucleon interaction. In this case, a polytrope can be used as an approximation for the neutron matter equation of state
\begin{equation}\label{eq:eos0}
    p=\kappa\rho^{1+1/n},
\end{equation}
where $\kappa$ is a constant and $n$ the so-called polytropic index. Note that most realistic equations of  state for neutron stars can be approximated as a polytrope or compound polytropes with different effective indices in the range $n\simeq 0.5-1.0$. It is also pointed that, in gravitational units, $K^{n/2}$ has units of length and so it can be used to set the fundamental length scale of the system \citep{1994ApJ...422..227C}. In our numerical computations, we will use the following dimensionless quantities ${\rho}_{\ast} = \kappa^n \rho$, ${r}_{\ast} = \kappa^{-n/2} r$, $m_{\ast}= \kappa^{-n/2}\, m$, etc. 

At the center of the star ($r = 0$), because of  the finiteness of the central pressure and energy density (i.e., $p(r=0) = p_0$ and $\rho(r=0) = \rho_0$ are two constants),  we expect  the enclosed mass to be zero, $m(r = 0) =0$.    Near the center of the star, it can be obtained that the asymptotic forms  of $m$ and $p$ read 
\begin{equation}
    m(r) = m_3 r^3 + m_5 r^5 +\mathcal{O}(r^7)
\end{equation}  
and 
\begin{equation} 
    p(r) = p_0 + p_2 r^2 +\mathcal{O}(r^4),
\end{equation}
respectively, where 
\begin{eqnarray}
    m_3 & = & \frac{2 \pi  \left[(3 \eta +2) \rho _0+3 \eta  p_0\right]}{3 (2 \eta +1)}, \\ 
    m_5 & = & \frac{2 \pi  \left(3 c_{s0}^2 \eta +3 \eta +2\right) p_2}{5 (2 \eta +1)c_{s0}^2 } ,\\ 
    p_2 & = & -\frac{4 \pi c_{s0}^2 \left(p_0+\rho _0\right) \left[(3 \eta +1) \rho _0+3 (\eta +1) p_0\right]}{3 \left[c_{s0}^2 (\eta +2)+\eta \right]} \nonumber\\
\end{eqnarray}
and  $c_{s0}^2$ is the value of speed of sound at the center of the star. Clearly, when $\eta =0$, the above asymptotic forms of $m$ and $p$ go back to the same as those in GR \citep{abdelsalhin2019tidal}. 

To obtain an equilibrium configuration of the star with total mass $M=m(R)$, we need integrating Eqs. (\ref{eq:TOVm}) and (\ref{eq:TOVp}), along with the equation of state (\ref{eq:eos0}), from the center to its surface $r=R$, where $R$ denotes the radius of the star. The internal solution should be matched to the external one that is also a Schwarzschild solution as in GR.  Therefore, on the surface of the star, the pressure as well as the energy density vanish, i.e. $p(R)=0$, $\rho(R) = 0$. 

\section{Perturbations and tidal Love numbers}\label{sec:TLN}
Now, let us consider a static neutron star immersed in a weak external tidal field. We further assume that  the time dependence of the tidal field can be  neglected.  In this circumstance, the star will be deformed and develope  a multipolar structure in response to the tidal field. Then, the metric of the spacetime outside the star will be accordingly deformed. 

We express the metric of the  deformed spacetime as
\begin{equation}
    g_{\mu\nu} = g_{\mu\nu}^{(0)} +h_{\mu\nu},
\end{equation}
where $h_{\mu\nu}$ is a small perturbation to the background metric $g_{\mu\nu}^{(0)}$ which is given by integrating generalized TOV equations (\ref{eq:TOVm}), (\ref{eq:TOVp}) and (\ref{eq:TOValpha}).  

The perturbation $h_{\mu\nu}$ can be decomposed in spherical harmonics and separated into even  and odd  parts 
\begin{equation}
    h_{\mu\nu} = h_{\mu\nu}^{\mathrm{even}} +h_{\mu\nu}^{\mathrm{odd}},
\end{equation}
according to  parity under the rotation in ($\theta, \phi$) plane.
In Regge-Wheeler gauge, the two part of $h_{\mu\nu}$ can be respectively written as \citep{Binnington:2009bb,Cardoso2017}
\begin{equation}\label{eq:h_even}
    h_{\mu \nu}^{\mathrm{even}}=\left(\begin{array}{cccc}
        -e^{2 \alpha} H_{0} & H_{1} & 0 & 0 \\
        H_{1} & e^{2 \beta} H_{2} & 0 & 0 \\
        0 & 0 & K r^{2} & 0 \\
        0 & 0 & 0 & K r^{2} \sin ^{2} \theta
        \end{array}\right) Y^{lm}
\end{equation}
and
\begin{equation}\label{eq:h_odd}
    h_{\mu \nu}^{\mathrm{odd}}=\left(\begin{array}{cccc}
        0 & 0 & h_0 S_{\theta}^{lm} & h_0 S_{\phi}^{lm} \\
       0 & 0 & h_1 S_{\theta}^{lm} &  h_1 S_{\phi}^{lm} \\
       h_0 S_{\theta}^{lm} & h_1 S_{\theta}^{lm} & 0 & 0 \\
       h_0 S_{\phi}^{lm} &  h_1 S_{\phi}^{lm} & 0 & 0
        \end{array}\right),
\end{equation}
where $H_0$, $H_1$, $H_2$, $K$, $h_0$ and $h_1$ are functions of $r$. Here, $Y^{lm}$ is the scalar harmonics and $S_{\theta} \equiv -Y_{,\phi}^{lm}/\sin\theta$ and $ S_{\phi} \equiv \sin\theta Y_{,\theta}^{lm}$ \citep{abdelsalhin2019tidal}.

Spacetime fluctuations are accompanied by the corresponding fluctuations in matter field. Because the density $\rho$ and pressure $p$ of the fluid in star are both scalar fields, the components of the  perturbations of the energy-momentum tensor (\ref{eq:EMT})  read
\begin{equation}\label{eq:deltaTmunu}
    \delta T_{\mu}^{\nu} = \mathrm{diag}(-\delta \rho, \delta p, \delta p, \delta p) Y^{lm}.
\end{equation}
To the linear level, the even and odd sectors of the perturbations are independent to each other, therefore, we can discuss them individually.

\subsection{Even-parity sector}
Inserting the perturbation quantities in Eq.(\ref{eq:h_even}) and Eq.(\ref{eq:deltaTmunu}) into the linearized Einstein equations 
\begin{equation}\label{eq:linearizedEinsteinEq}
    \delta R_{\mu}^{\nu} = 8 \pi \left(\delta T_{\mu}^{\nu}-\frac{1+\eta}{1+2\eta}\delta^{\nu}_{\mu} \frac{\delta T}{2}\right),
\end{equation} 
where $\delta T$ is trace of $\delta T_{\mu}^{\nu}$, we find that $H_2=H_0$, $H_1=0$,  and $K$ can be written as a function of $H_0$ and $\delta p$, whereas $H_0$ obeys  the following homogeneous differential equation
\begin{equation}\label{eq-H_0}
    H_0''+ P(r) H_0' + Q(r) H_0=0,
\end{equation} 
where 
\begin{equation}
    P(r) = \frac{2}{r}+e^{2\beta}\left(\frac{2m}{r^2}-\frac{4\pi r[(\rho-p)+\eta(\rho+p)]}{1+2\eta}\right),
\end{equation}
\begin{eqnarray}
    Q(r) &=  &  8 \pi  e^{2 \beta } \left( \frac{\left[5 \eta  ({c_s^2}+1)+(7 {c_s^2}+1)\right] (p+\rho)}{\eta  ({c_s^2}+1)+2 {c_s^2}}\right.\nonumber\\ 
     & -& \hspace{-2mm}\left.\frac{\eta  (p+\rho )+(\rho-p)}{1+ 2 \eta}\right)-\frac{ l (l+1)}{r^2}e^{2 \beta }\nonumber \\ &-& 4\left(\frac{d\alpha}{dr}\right)^2.
\end{eqnarray}

Outside the star, $\rho = p = 0$ and the mass function $m(r)$ takes a constant value that is equal to the mass of the star $M$. Therefore, Eq.(\ref{eq-H_0}) reduces to
\begin{eqnarray}\label{eq:H0}
    H_0''&+&\frac{2(r-M)}{r(r-2M)}H_0'\\ 
    &-&\frac{4M^2-2l(l+1)Mr+l(l+1)r^2}{r^2(r-2M)^2}H_0=0.\nonumber
\end{eqnarray} 
The asymptotic behaviour of the solution of above equation at large $r$ reads
\begin{equation}
    H_0=c_1^{\infty}{r^{-(l+1)}} + \mathcal{O}(r^{-(l+2)})+ c_2^{\infty}r^l+\mathcal{O}(r^{l-1}),
\end{equation}
where two constants $c_1^{\infty}$ and $c_2^{\infty}$  are related to the  electric-type tidal Love numbers $k_l^E$ via 
\begin{equation}
    k_2^{E} = \frac{1}{2R^{2l+1}}\frac{c_1^{\infty}}{c_2^{\infty}}.
\end{equation}

In fact, the general solution of Eq.(\ref{eq:H0}) can be exactly expressed in terms of the associated Legendre functions $Q_l^2$ and $P_l^2$, i.e. 
\begin{equation}
    H_0= c_1 Q_l^2\left(\frac{r}{M}-1\right)+c_2 P_l^2\left(\frac{r}{M}-1\right),
\end{equation} 
where $c_1$ and $c_2$ are two constants to be determined by matching  interior solution of $H_0$ at the stellar surface.

Inside the star, we require $H_0$ is regular at the center $r=0$. Hence, from the asymptotic form of Eq.(\ref{eq-H_0}) near $r = 0$, we have $H_0 \sim r^{l}$. Taking this as the initial condition, we can integrate Eq.(\ref{eq-H_0}) numerically to find the value of $H_0$ and $H_0'$ at the surface $r = R$. 

\subsection{Odd-parity sector}

In this situation, since there exists no odd-parity matter  in the model,  from Eq.(\ref{eq:linearizedEinsteinEq}), it is easily obtained that $h_1 =0$, and $h_0$ obeys 
\begin{equation}\label{eq-h_0}
    h_0''+\tilde{P}(r)h_0'+\tilde{Q}(r)h_0 =0,
\end{equation}
where
\begin{equation}
    \tilde{P}(r)=-\frac{4\pi r^2}{r-2m}(\rho+p),
\end{equation}
\begin{equation}
    \tilde{Q}(r)=\frac{4m-l(l+1)r}{r^2(r-2m)} +\frac{8\pi r\left(\rho-p +\eta(\rho+p)\right)}{(1+2\eta)(r-2m)}.
\end{equation}
Outside the star, Eq.(\ref{eq-h_0}) reduces to 
\begin{equation}
    h_0''+\frac{4M-l(l+1)r}{r^2(r-2M)}h_0=0.
\end{equation}
Similar to the even sector, for $r \to \infty$, the exterior solution of Eq.(\ref{eq-h_0})  can be asymptotically expressed as
\begin{equation}
    h_0=\tilde{c}_1^{\infty}{r^{-l}} + \mathcal{O}(r^{-l-1})+ \tilde{c}_2^{\infty}r^{l+1}+\mathcal{O}(r^{l}),
\end{equation}
from which the so-called magnetic-type tidal Love numbers are obtained by 
\begin{equation}
    k_2^{M} = -\frac{l}{2(l+1)}\frac{1}{R^{2l+1}}\frac{\tilde{c}_1^{\infty}}{\tilde{c}_2^{\infty}}.
\end{equation}

\subsection{Quadrupolar tidal Love numbers}

The quadrupolar ($l = 2$) perturbations give the main contribution to the stellar deformations. Here, we focus our analysis on the  quadrupolar case.

In the even-parity sector, as $r \to \infty$, 
\begin{equation}
    H_0 = \frac{8}{5}c_1\left(\frac{M}{r}\right)^3 +\mathcal{O}\left(\frac{M}{r}\right)^4+ 3c_2\left(\frac{r}{M}\right)^2 +\mathcal{O}\left(\frac{r}{M}\right),
\end{equation}
that is, $c_1^{\infty} = \frac{8}{5}c_1 M^3$ and $c_2^{\infty} = 3c_2M^{-2}$.
Therefore, using the relation 
\begin{equation}
    \frac{c_1}{c_2} = \frac{r \frac{\partial }{\partial r}P_{2 }^2\left(\frac{r}{M}-1\right)-y P_{2 }^2\left(\frac{r}{M}-1\right)}{y Q_{2 }^2\left(\frac{r}{M}-1\right)-r \frac{\partial }{\partial r}Q_{2 }^2\left(\frac{r}{M}-1\right)},
\end{equation}
where $y\equiv R H_0'(R)/H_0(R)$,  we have
\begin{equation}\label{eq:k2e}
    \begin{aligned}
    k_{2}^E 
    &= \frac{8 C^{5}}{5}(1-2 C)^{2}[2+2 C(y-1)-y] \\
    & \times\left\{2 C(6-3 y+3 C(5 y-8))\right.\\ &+4 C^{3}\left[13-11 y+C(3 y-2)+2 C^{2}(1+y)\right]\\
    &\left.+3(1-2 C)^{2}[2-y+2 C(y-1)] \log (1-2 C)\right\}^{-1},
    \end{aligned}
\end{equation}
where $C \equiv M/R$ denoting the compactness of the star. 

Similarly, In the odd-parity sector, 
\begin{equation}
    h_0=-\frac{1}{5} c_1 \left(\frac{M}{r}\right)^2+\mathcal{O}\left(\frac{M}{r}\right)^3+ c_2\left(\frac{r}{M}\right)^3 +\mathcal{O}\left(\frac{r}{M}\right)^2
\end{equation}
for $r\to \infty$, then the  magnetic-type TLN can be expressed as
\begin{equation}\label{eq:k2m}
    \begin{aligned}
        k_{2}^M &= \frac{8 C^{5}}{5} [2 C (y-2)-y+3] \\
        & \times\left\{2 C \left[2 C^3 (y+1)+2 C^2 y+3 C (y-1)-3 y+9\right]\right.\\
        &\left.+3 [2 C (y-2)-y+3] \log (1-2 C)\right\}^{-1}.
        \end{aligned} 
\end{equation}
Note that the formula Eqs.(\ref{eq:k2e}) and (\ref{eq:k2m}) are the same as those obtained in GR\citep{Hinderer:2007mb,Cardoso2017,Lau_2019}. This is reasonable due to the fact that outside the star is a vacuum in which Rastall gravity is equivalent to GR.

\begin{figure}[ht]
    \includegraphics[width=\columnwidth]{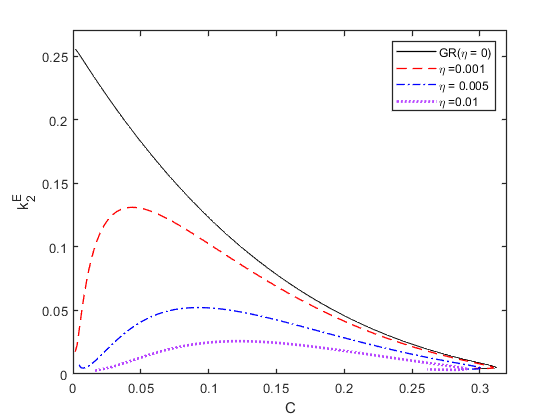}
    \caption{The electric-type quadrupolar TLNs $k_2^E$ as a function of the neutron star's compactness $C$ for different values of the parameter $\eta$, where the value of polytropic index $n =1$ is taken. }
   \end{figure}

   \begin{figure}[ht]
    \includegraphics[width=\columnwidth]{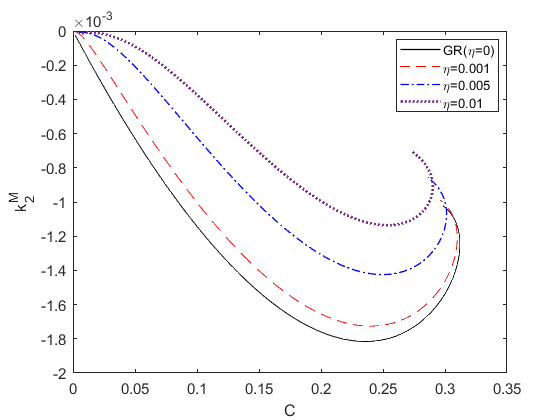}
    \caption{The magnetic-type quadrupolar TLNs $k_2^M$ as a function of the neutron star's compactness $C$ for different values of the parameter $\eta$, where the value of polytropic index $n =1$ is taken. }
   \end{figure}

However, the value of the quantity $y$ in Eqs.(\ref{eq:k2e}) and (\ref{eq:k2m}) should be respectively obtained from Eqs.(\ref{eq-H_0}) and (\ref{eq-h_0}) by integrating them  in the range $0<r<R$, which are expected to be different from those in GR. This point is confirmed by our numerical results.  In Figs. 1 and 2, we plot $k_2^E$ and $k_2^M$  as functions of the neutron star's compactness $C$, respectively, for some different values of the parameter $\eta$.  Compared to those in GR, it is obviously that both the electric-type and magnetic-type TLNs of the star are smaller than those in GR for the same value of compactness, which means neutron stars in Rastall gravity is harder to deform in the external tidal field. For a less compact neutron star, the electric-type TLNs are heavily depressed in Rastall gravity, whereas the magnetic-type ones are close to those in GR. On the contrary, for a high compact star, the magnetic-type TLNs are significantly smaller than those in GR, but the electric-type ones become close to those in GR.

\section{Discussions}
In summary, we have computed both of the two types of TLNs of a compact star with polytrope matter in Rastall gravity. It is interesting to work in models with a more realistic EOS and compare them with the constraints coming from the observation of the TLNs of merging neutron stars \citep{LIGOScientific:2017vwq}.

From our numerical results, the value of TLNs are heavily depressed by the deviation parameter $\eta$. It seems that this depression effect for the TLNs can be tested by current detections. However, this effect is highly degenerated with that from EOS of nuclear matter in neutron star. In a model with more realistic EOS, there may exists similar depression, see Ref.\citep{abdelsalhin2019tidal} for example.  Due to the large uncertainty about the EOS of the matter in compact stars, the effect  has little measurable influence  at least on the current gravitational wave observations.  We expect near future observations and a better understanding on the ultradense nuclear matter can take the deviations to be measurable.

It is worth noting that in this work, the  parameter $\eta$ takes relatively small value. This is because the $\eta$ is also constrained by other experiments, such as local measurements of the gravitational constant \citep{Rosi:2014kva}. Besides, because of the divergence of the numerical evaluation for negative $\eta $, we do not take this circumstance into account which deserves further investigations.

\acknowledgments
The authors are indebted to Ping Xi for helpful discussions.

\begin{fundinginformation}
    This research was supported by innovation program of Shanghai Normal University under Grant No. KF202147.
    \end{fundinginformation}

    \begin{conflict}
    The authors declare that they have no conflicts of interest.
    \end{conflict}

\bibliographystyle{spr-mp-nameyear-cnd}  
\bibliography{refs}                

\end{document}